\documentclass[prd,twocolumn,showpacs,floatfix,amsmath,amssymb,nofootinbib]{revtex4}

\usepackage{amsmath,amssymb,mathrsfs,bm}
\usepackage{cases}

\usepackage{graphicx}
\usepackage{overpic}
\usepackage{booktabs}
\usepackage{multirow}
\usepackage{array}

\usepackage{color}
\usepackage{indentfirst}

\usepackage[colorlinks,
citecolor=blue,
linkcolor=red,
urlcolor=blue]{hyperref}
\usepackage{cleveref}

\usepackage{enumitem}
\newcolumntype{C}[1]{>{\centering\arraybackslash}p{#1}}

\begin{document}
    
\title{Spectroscopy and femtoscopic correlation function of the $B\bar{D}$, $B=(N, \Delta)$ system in quark delocalization color screening model }

\author{Xuejie Liu$^1$}\email[E-mail: ]{1830592517@qq.com}
\author{Dianyong Chen$^{2,5}$\footnote{Corresponding author}}\email[E-mail:]{chendy@seu.edu.cn}
\author{Hongxia Huang$^4$}
\email[E-mail: ]{hxhuang@njnu.edu.cn}
\author{Jialun Ping$^4$}
\email[E-mail: ]{jlping@njnu.edu.cn}
\affiliation{$^1$School of Physics, Henan Normal University, Xinxiang 453007, P. R. China}
\affiliation{$^2$School of Physics, Southeast University, Nanjing 210094, P. R. China}
\affiliation{$^4$Department of Physics, Nanjing Normal University, Nanjing 210023, P.R. China}
\affiliation{$^5$Lanzhou Center for Theoretical Physics, Lanzhou University, Lanzhou 730000, P. R. China}

\begin{abstract}
  In this work, we systematically investigate the pentaquark systems with quark contents $qqqq\bar{c}$ with the analyzed total spin and parity quantum numbers of  $J^{P}=\frac{1}{2}^{-}$, $J^{P}=\frac{3}{2}^{-}$ and $J^{P}=\frac{5}{2}^{-}$, in the I=0, I=1 and I=2 isospin channels. The effective potentials between baryon and meson clusters are given, and the possible bound states are also investigated. Also, the study of the scattering process of the open channels is performed to identify possible resonance states. Our estimations indicate that several possible bound states and narrow baryon-meson resonances are found from corresponding the calculation processes. Furthermore, to bridge the gap between theoretical predictions and experimental measurement, we also extract the low-energy scattering parameters and compute the femtoscopic correlation functions for the $N\bar{D}$ system using the CATS framework. The results demonstrate that the predicted $I=0$ bound state manifests as a significant enhancement at low momentum accompanied by a characteristic suppression. In contrast, the $I=1$ correlations remain relatively flat as the predicted resonances are kinematically distant from the threshold. The $I=2$ sector exhibits strong spin dependence, where the bound state signal in the $J=1/2^{-}$ channel is largely masked by repulsive components in spin-averaged observables. This cancellation effect suggests that future experimental searches at  ALICE and LHCb may require spin-selective measurements to identify such states.  These predictions provide crucial theoretical guidance for future experiments.
\end{abstract}

\maketitle

\setcounter{totalnumber}{5}
\section{\label{sec:introduction}Introduction}
Hadron physics has opened a renewed interest in multi-hadron system. In the past years, many new hadronic states have been observed experimentally, and they carry exotic quantum numbers that cannot be reached by the standard quark model for meson of $q\bar{q}$ and for baryons for $qqq$. This significant progress in experiments has triggered plenty of theoretical interest and made the study of these exotic states an intriguing topic in hadronic physics~\cite{Klempt:2007cp,Brambilla:2010cs,Hosaka:2016pey,Chen:2016qju,Lebed:2016hpi,Esposito:2016noz,Dong:2017gaw,Ali:2017jda,Guo:2017jvc,Olsen:2017bmm,Karliner:2017qhf,Liu:2019zoy,
Brambilla:2019esw,Richard:2019cmi}. Among these states, one can highlight the new three hidden charm pentaquark candidates observed in 2019 by the LHCb collaboration~\cite{LHCb:2019kea} in the $J/\psi p$ invariant mass spectrum of $\Lambda_{b}^{0}\longrightarrow J/\psi K^-p$ decays; they were defined as $P_{c}(4312)$, $P_{c}(4440)$ and $P_{c}(4457)$, respectively. In fact, the hidden charm pentaquark could be tracked back to 2015, when two exotic states: $P_{c}(4380)$ and $P_{c}(4450)$ were be found in the same decay process at the same collaboration~\cite{LHCb:2015yax}. In principle, due to the fact that the $P_{c}$ states were found in the experiments, the $P_{cs}$ states as their strangeness partners should also exist. So, recently, the LHCb Collaboration released their results about the $\Xi_{b}\rightarrow J/\psi K^-\Lambda$ decay, which indicates a new resonance structure named $P_{cs}(4459)$. Besides the exotic states mentioned above, , the existence of flavor-exotic states, where quarks and antiquarks can not annihilate through strong and electromagnetic interactions, was also predicted. Therefore, searching for these flavor-exotic states has become increasingly important both theoretically and experimentally.

The discussion about the nature of these exotic signals are carried out by various theoretical approaches. For instance, the three newly announced hidden-charm pentaquarks, $P_{c}(4312)$, $P_{c}(4440)$ and $P_{c}(4457)$ are more likely to be considered to be the molecular state of $\Sigma_{c}\bar{D}^{*}$ in effective field theories~\cite{Liu:2019tjn,He:2019ify}, QCD sum rules~\cite{Wang:2019got}, phenomenological potential models~\cite{Guo:2019kdc,Huang:2018wed,Mutuk:2019snd,Zhu:2019iwm,Eides:2019tgv,Weng:2019ynv}, heavy quark spin symmetry~\cite{Shimizu:2019ptd,Xiao:2019aya} and heavy hadron chiral perturbation theory ~\cite{Meng:2019ilv}. Moreover, their photo-production ~\cite{Cao:2019kst,Wang:2019krd} and decay properties~\cite{Xiao:2019mvs} have been also discussed. As for the other types of pentaquarks, which are genuinely exotic states with no quark-antiquark annihilation although the existence of several states has not been experimentally confirmed. For example, the pentaquarks state composed of $uudu\bar{s}$ triggered a lot of theoretical work on mutli-quark system. In Ref.~\cite{Hiyama:2005cf}, the authors used the standard nonrelativistic quark model of Isgur-Karl to investigate the NK scattering problem, and the NK scattering phase shift showed no resonance in the energy region 0-500 MeV above the NK threshold. In Refs.~\cite{Lemaire:2003et,Liu:2019sip}, the NK phase shifts were calculated within a constituent quark model by numerically solving the RGM equation. Wang et al.~\cite{Wang:2004ky} gave a study on the NK elastic scattering in a quark potential model and their results are consistent with the experimental data. In Ref.~\cite{Barnes:1992ca}, Barnes and Swanson used the quark-Born-diagram method to derive the NK scattering amplitudes and obtained reasonable results for the NK phase shifts, but they were limited to the $S-$wave. In Ref.~\cite{Huang:2004sj}, the NK interaction was studied in the constituent quark model and the numerical results of different partial waves were in good agreement with the experimental data.

 Now the pentaquarks composed of $uudu\bar{s}$ extends to the heavy quark sector, there have been, not only analogous discussions, but also new approaches which are not accessible in light flavor sectors. There, an interesting observation is made that there is a sufficiently strong attraction between a $\bar{D}(\bar{D}^{*})$ meson and a nucleon N. As a matter of fact, many theoretical studies have been devoted to the $N\bar{D}$ system. Since the hadron nucleon interaction is the basic quantity for the hadronic molecules and for exotic nuclei, the interaction of $N\bar{D}$ system is investigated, which is subtle~\cite{Yasui:2009bz,Yamaguchi:2011xb,Gamermann:2010zz,Haidenbauer:2007jq,Carames:2012bd,Hosaka:2016ypm}. Bound states of nucleon and an open heavy meson are discussed with respect to the heavy quark symmetry~\cite{Yasui:2009bz}, and the results that indicated the $N\bar{D}$ with $IJ^{P}=0\frac{1}{2}^-$ is a bound state.  In Ref.~\cite{Yamaguchi:2011xb}, the bound state found previously in the $I(J^{P})=0(\frac{1}{2}^-)$ channel survives when short range interaction is included and  a resonance state with $I(J^{P})=0(\frac{3}{2}^-)$ as a Feshbach resonance predominated by a heavy vector meson and a nucleon($N\bar{D}^{\ast}$). In heavy quark symmetry its result implied that the coupling to $N\bar{D}^{\ast}$ channel bring to attraction to the $N\bar{D}$ interaction and generates an $S-$wave $N\bar{D}$ bound state with binding energy about 1 MeV~\cite{Gamermann:2010zz}. However, some models suggested no significant attraction in the $N\bar{D}$ interaction~\cite{Haidenbauer:2007jq,Carames:2012bd}. Hence, it is worthwhile to make a systematical study of the $N\bar{D}$ system by using different methods, which will deepen our understanding of possible pentaquarks.

 Given the limited experimental data in the heavy flavor sector , theoretical insights from proven models are crucial, especially to interpret recent proton-charm femtoscopy measurements by ALICE collaboration~\cite{ALICE:2022enj}. In this work,  we utilize the quark delocalization color screening model (QDCSM)—consistent with the LHCb $P_{c}$ states~\cite{LHCb:2019kea}—to study $qqqq\bar{c}$ pentaquarks with quantum numbers $J^{P}=\frac{1}{2}^{-}$, $J^{P}=\frac{3}{2}^{-}$, and $J^{P}=\frac{5}{2}^{-}$, and in the I=0, 1 and 2 isospin sectors via the  resonating group method (RGM). Furthermore, we extend our investigation beyond spectroscopy to scattering observables. By extracting the scattering length and effective range from the RGM phase shifts, we construct equivalent potentials to calculate femtoscopic correlation functions.  Unlike previous studies that were limited to spectroscopy or phase shifts analysis. This comprehensive framework allows us to translate microscopic bound state predictions into specific correlation signals, providing a direct reference for the ongoing proton-charm femtoscopy measurements.

The plan of this paper is the following. In the next section the QDCSM are briefly presented. Section~\ref{dis} discusses the results of the effective potential, the bound state calculation, and the scattering phase shifts. In addition, the low-energy scattering parameters and the femtoscopic correlation functions are calculated and analyzed in detail. Section.~\ref{sum} is devoted to the summary and concluding remarks of this study.

\section{THE QUARK DELOCALIZATION COLOR SCREENING MODEL (QDCSM) \label{mod}}
In this work, the main purpose is to detect the presence of possible bound states or resonance states in the $B\bar{D}$ system. Now, we use the quark delocalization color screening model to calculate the spectra of the the $B\bar{D}$ system. Besides, we employ the resonating group (RGM) method to calculate the baryon-meson scattering phase shifts and to look for the resonance states. 

The quark delocalization color screening model (QDCSM) is an extension of the native quark cluster model~\cite{DeRujula:1975qlm,Isgur:1979be,Isgur:1978wd,Isgur:1978xj} and was developed with aim of
addressing multiquark systems. The detail of QDCSM can be found in the Refs.~\cite{Wang:1992wi,Chen:2007qn,Chen:2011zzb,Wu:1996fm,Huang:2011kf}.
Here, the general form of the five-body complex Hamiltonian is given by
\begin{equation}
H = \sum_{i=1}^{5} \left(m_i+\frac{\boldsymbol{p}_i^2}{2m_i}\right)-T_{CM}+\sum_{j>i=1}^5V(r_{ij}),\\
\end{equation}
where the center-of-mass kinetic energy, $T_{CM}$ is subtracted without losing generality since we mainly focus on the internal relative motions of the multiquark system. The interplay is of two body potential which includes color-confining, $V_{CON}$, one-gluon exchange, $V_{OGE}$, and Goldstone-boson exchange, $V_{\chi}$, respectively,
\begin{equation}
V(r_{ij}) = V_{CON}(r_{ij})+V_{OGE}(r_{ij})+V_{\chi}(r_{ij}).
\end{equation}

Note herein that the potential could contain central, spin-spin, spin-orbit, and tensor contributions; In this work, only the first two will be considered attending the goal of the present calculation and for clarity in our discussion.
The potential $V_{OGE}(r_{ij})$ can be written as
\begin{eqnarray*}
V_{OGE}(r_{ij}) &=& \frac{1}{4}\alpha_s \boldsymbol{\lambda}^{c}_i \cdot\boldsymbol{\lambda}^{c}_j \\
&&\left[\frac{1}{r_{ij}}-\frac{\pi}{2}\delta(\boldsymbol{r}_{ij})\left(\frac{1}{m^2_i}+\frac{1}{m^2_j}
+\frac{4\boldsymbol{\sigma}_i\cdot\boldsymbol{\sigma}_j}{3m_im_j}\right)\right],
\end{eqnarray*}
where $m_{i}$ and $\boldsymbol{\sigma}$ are the quark mass and the Pauli matrices, respectively. The $\boldsymbol{\lambda^{c}}$ is SU(3) color matrix. The QCD-inspired effective scale-dependent strong coupling constant, $\alpha_s$, offers a consistent description of mesons from light to heavy quark sector, which can be written by,

\begin{equation}
  \alpha_{s}(\mu)=\frac{\alpha_{0}}{\ln\left(\frac{\mu^2+\mu_{0}^{2}}{\Lambda_{0}^{2}}\right)}.
\end{equation}

Similary, the confining interaction $V_{CON}(r_{ij})$ can be expressed as
\begin{equation}
 V_{CON}(r_{ij}) =  -a_{c}\boldsymbol{\lambda^{c}_{i}\cdot\lambda^{c}_{j}}\left[f(r_{ij})+V_{0_{ij}}\right],
\end{equation}
and the $f(r_{ij})$ can be written as
\begin{equation}
 f(r_{ij}) =  \left\{ \begin{array}{ll}r_{ij}^2 &\qquad \mbox{if }i,j\mbox{ occur in the same cluster}, \\
\frac{1 - e^{-\mu_{ij} r_{ij}^2} }{\mu_{ij}} & \qquad \mbox{if }i,j\mbox{ occur in different cluster}, \\
\end{array} \right.
\end{equation}
where the color screening parameter $\mu_{ij}$ is determined by fitting the deuteron properties, $NN$ and $NY$ scattering phase shifts~\cite{Chen:2011zzb,Ping:1993me,Wang:1998nk}., with $\mu_{qq}= 0.45$, $\mu_{qs}= 0.19$
and $\mu_{ss}= 0.08$, satisfying the relation $\mu_{qs}^{2}=\mu_{qq}\mu_{ss}$, where $q$ represents $u$ or $d$ quark. When extending to the heavy-quark case, we found that the dependence of the parameter $\mu_{cc}$ is not very significant in the calculation of the $P_{c}$ states~\cite{Huang:2015uda} by taking it from $10^{-4}$ to $10^{-2}\ \mathrm{fm}^{-2}$. The typical size of the multiquark system is of several femtometres, thus the value of the $\mu_{ij} r^2$ is rather small, and in this case the exponential function can be approximated to be
\begin{eqnarray}\label{muij}
  e^{-\mu_{ij}r_{ij}^{2}} &=& 1-\mu_{ij}r_{ij}^{2}+\mathcal{O}\left(\mu_{ij}^2 r_{ij}^4\right).
\end{eqnarray}
Accordingly, the confinement potential between two clusters is approximated to be
\begin{eqnarray}
  V_{CON}(r_{ij}) &=&  -a_{c}\boldsymbol{\mathbf{\lambda}}^c_{i}\cdot
\boldsymbol{\mathbf{
\lambda}}^c_{j}~\left(\frac{1-e^{-\mu_{ij}\mathbf{r}_{ij}^2}}{\mu_{ij}}+
V_{0_{ij}}\right) \nonumber \\
  ~ &\approx & -a_{c}\boldsymbol{\mathbf{\lambda}}^c_{i}\cdot
\boldsymbol{\mathbf{ \lambda}}^c_{j}~\left(r_{ij}^2+ V_{0_{ij}}\right),
\end{eqnarray}
which is the same with the expession of two quarks in the same cluster. Thus, when the value of the $\mu_{ij}$ is very small, the screened confinement will return to the quadratic form, which is why the results are insensitive to the value of $\mu_{cc}$. In the present work, we take $\mu_{cc}=0.01$. Then $\mu_{sc}$ and $\mu_{uc}$ are obtained by the relation $\mu_{sc}^{2}=\mu_{ss}\mu_{cc} $ and $\mu_{uc}^{2}=\mu_{uu}\mu_{cc}$, respectively. 

The Goldstone-boson exchange interactions between light quarks appear because the dynamical breaking of chiral symmetry. For the $B\bar{D}$ system, the $K$ exchange interactions do not appear because there is no $s$ quark herein. Only the following $\pi$ and $\eta$ exchange term works between the chiral quark-(anti)quark pair.
\begin{eqnarray*}
V_{\chi}(r_{ij}) & =&  v^{\pi}_{ij}(r_{ij})\sum_{a=1}^{3}\lambda_{i}^{a}\lambda_{j}^{a}+v^{\eta}_{ij}(r_{ij})\\
&&\left[\left(\lambda _{i}^{8}\cdot
\lambda _{j}^{8}\right)\cos\theta_P-\left(\lambda _{i}^{0}\cdot
\lambda_{j}^{0}\right) \sin\theta_P\right] \label{sala-Vchi1}
\end{eqnarray*}
with
\begin{eqnarray*}
  v^{B}_{ij} &=&  {\frac{g_{ch}^{2}}{{4\pi}}}{\frac{m_{B}^{2}}{{\
12m_{i}m_{j}}}}{\frac{\Lambda _{B}^{2}}{{\Lambda _{B}^{2}-m_{B}^{2}}}}
m_{B}         \\
&&\left\{(\boldsymbol{\sigma}_{i}\cdot\boldsymbol{\sigma}_{j})
\left[ Y(m_{B}\,r_{ij})-{\frac{\Lambda_{B}^{3}}{m_{B}^{3}}}
Y(\Lambda _{B}\,r_{ij})\right] \right\},
B=\pi, \eta,
\end{eqnarray*}

where $Y(x)=e^{-x}/x$ is the standard Yukawa function. The $\boldsymbol{\lambda^{a}}$ is the SU(3) flavor Gell-Mann matrix. The mass of the $\eta$ and $\pi$ meson is taken from the experimental value~\cite{ParticleDataGroup:2018ovx}. Finally, the chair coupling constant, $g_{ch}$, is determined from the $\pi NN$ coupling constant through~\cite{Vijande:2004he,Fernandez:1986zn}
\begin{equation}
\frac{g_{ch}^{2}}{4\pi}=\left(\frac{3}{5}\right)^{2} \frac{g_{\pi NN}^{2}}{4\pi} {\frac{m_{u,d}^{2}}{m_{N}^{2}}}
\end{equation}
which assumes that flavor SU(3) is an exact symmetry, only broken by the different mass of the strange quark. The model parameters and the masses of the ground mesons are listed in Tables~\ref{parameters} and \ref{mass}, respectively.

\begin{table}[ht]
\caption{\label{biaoge}The values of the Model parameters. The masses of mesons take their experimental values.}
\renewcommand\arraystretch{1.3}
\begin{tabular}{p{2.5cm}<\centering p{2.5cm}<\centering p{2.5cm}<\centering }
 \toprule[1pt]
      & Parameter  &Value   \\
      \midrule[1pt]
Quark masses  &$m_u(MeV)$                    & 313 \\
              &$m_s(MeV)$                      & 573 \\
              &$m_c(MeV)$                      & 1675 \\
              &$m_b(MeV)$                      & 5086 \\
      \midrule[0.5pt]
confinement   &$b(fm)$                          &0.518\\
              &$a_{c}$(MeV $fm^{-2}$)           &58.03 \\
              &$V_{0}(fm^2)$                    &-1.2883\\
     \midrule[0.5pt]
OGE           &$\alpha_{0}$                     &0.5101\\
              &$\Lambda_{0}(fm^{-1})$           &1.525\\
              &$\mu_{0}(MeV)$                   &445.808\\
              &$\alpha_{ch}$                    &0.027\\
              &$\mu_{uu}(fm^{-2})$              &0.45\\
              &$\mu_{ss}(fm^{-2})$              &0.08\\
    \midrule[0.5pt]
Goldstone boson     & $m_\pi(fm^{-1})$ & 0.7 \\
                    & $m_\eta(fm^{-1})$  & 2.77\\
                    & $\Lambda_{\pi}(fm^{-1})$ &4.2\\
                    & $\Lambda_{\eta }(fm^{-1})$    &5.2\\

\bottomrule[1pt]
\end{tabular}
\label{parameters}
\end{table}

\begin{table}[ht]
\caption{The Masses (in MeV) of the ground baryons and mesons. Experimental values are taken
from the Particle Data Group (PDG)~\cite{ParticleDataGroup:2018ovx}.}
\renewcommand\arraystretch{1.2}
\begin{tabular}{p{1.5cm}<\centering p{1cm}<\centering  p{1cm}<\centering p{1cm}<\centering p{1cm}<\centering p{1cm}<\centering p{1cm}<\centering }
\toprule[1pt]
~~~~~~&~~$N$~~  &~~$\Delta$~~  &~~$\Lambda$~~ &~~$\Sigma$~~ &~~$\Sigma^{*}$~~ &~~$\Xi$~~~~~~ \\ \midrule[1pt]
Experiment  &939      & 1232         &1124          &1238         &1360             &1374     \\
Model       &939      & 1232         &1116          &1193         &1385             &1318     \\ \midrule[1pt]
~~~~~~&~~$\Xi^{*}$~~  &~~$\Omega$~~  &~~$\bar{D}$~~ &~~$\bar{D}^{*}$~~ &~~$\bar{B}$~~ &~~$\bar{B}^{*}$~~~~~~ \\\midrule[1pt]
Experiment  &1496           &1642          &1865                 &1900       &5279          &5290         \\
Model       &1533           &1672          &1864                 &2007       &5279          &5325       \\
\bottomrule[1pt]
\end{tabular}
\label{mass}
\end{table}

In QDCSM, the quark delocalization is realized by specifying the single particle orbital
wave function of QDCSM as a linear combination of left and right Gaussian, the single
particle orbital wave functions used in the ordinary quark cluster model,
\begin{eqnarray}\label{wave0}
\psi_{\alpha}(\boldsymbol{s}_{i},\epsilon)&=&\left(\Phi_{\alpha}(\boldsymbol{s}_{i})
  +\epsilon\Phi_{\beta}(\boldsymbol{s}_{i})\right)/N(\epsilon), \nonumber \\
\psi_{\beta}(\boldsymbol{s}_{i},\epsilon)&=&\left(\Phi_{\beta}(\boldsymbol{s}_{i})
  +\epsilon\Phi_{\alpha}(\boldsymbol{s}_{i})\right)/N(\epsilon), \nonumber \\
N(\epsilon)&=& \sqrt{1+\epsilon^2+2\epsilon e^{-s^2_{i}/{4b^2}}},\nonumber \\
\Phi_{\alpha}(\boldsymbol{s}_{i})&=&\left(\frac{1}{\pi b^2}\right)^{\frac{3}{4}}
e^{-\frac{1}{2b^2}\left(\boldsymbol{r_\alpha}-\frac{2}{5}s_{i}\right)^2},\nonumber \\
\Phi_{\beta}(-\boldsymbol{s}_{i})&=&\left(\frac{1}{\pi b^2}\right)^{\frac{3}{4}}
e^{-\frac{1}{2b^2}\left(\boldsymbol{r_\beta}+\frac{3}{5}s_{i}\right)^2},
\end{eqnarray}
The $\boldsymbol{s}_{i}$, $i=1,2,..., n$, are the generating coordinates, which are introduced to
expand the relative motion wave function~\cite{Wu:1998wu,Ping:1998si,Pang:2001xx}. The mixing parameter
$\epsilon(s_{i})$ is not an adjusted one but determined variationally by the dynamics of the
multi-quark system itself. This assumption allows the multi-quark system to choose its
favorable configuration in the interacting process. It has been used to explain the cross-over
transition between the hadron phase and the quark-gluon plasma phase~\cite{Xu:2007oam}.

\section{The results and discussions\label{dis}}

 In the present calculation, we investigate the possible lowest-lying and resonance states of the $uudu\bar{c}$ pentaquark systems by taking into account the $(uud)(u\bar{c})$ configurations. For the $uudu\bar{c}$ pentaquark systems, the considered baryons are always positive parity and mesons are either pseudoscalars ($0^-$) or vectors ($1^{-}$). So a pentaquark state with negative parity has $L=0$. Accordingly, the total angular momentum, $J$, coincides with the total spin, S, and can take values $\frac{1}{2}, \frac{3}{2}$ and $\frac{5}{2}$. The possible baryon-meson channels which are under consideration in the computation are listed in Table ~\ref{channels}; they are grouped according to total spin and parity $J^{P}$, and isospin $I$. Our purpose is to explore if there is any other pentaquark state, and to see whether those pentaquark states can be explained as the molecular pentaquarks. Since the attractive potential is the necessary for forming bound states, for the first step,  the effective potential of all channels is studied. Second, a dynamic calculation including both the single-channel and channel-coupling is carried out in order to check whether there is any bound state. Third, the scattering process of the open channels is observed to search for any resonance states. Finally, to bridge the gap between theory and experiment, we extract the low-energy scattering parameters from the calculated phase shifts and investigate the femtoscopic correlation functions for the relevant channels.

\begin{table}[!htb]
\renewcommand{\arraystretch}{1.5}
\begin{center}
\caption{\label{channels} All possible channels for all quantum numbers. }
\begin{tabular}{p{1.6cm}<\centering p{1.6cm}<\centering p{1.2cm}<\centering p{1.2cm}<\centering p{1.2cm}<\centering} 
\toprule[1pt]
$I=0$    &$S=\frac{3}{2}$    & N$\bar{D}^{*}$ &                \\
$I=1$    &$S=\frac{1}{2}$    & N$\bar{D}$     &N$\bar{D}^{*}$   &$\Delta\bar{D}^{*}$~~~~~~~~~\\
$I=1$    &$S=\frac{3}{2}$    & N$\bar{D}^{*}$ &$\Delta\bar{D}$  &$\Delta\bar{D}^{*}$~~~~~~~~~\\
$I=1$    &$S=\frac{5}{2}$    &$\Delta\bar{D}^{*}$ & &~~~~~~~~~\\
$I=2$    &$S=\frac{1}{2}$    &$\Delta\bar{D}^{*}$ & &~~~~~~~~~\\
$I=2$    &$S=\frac{3}{2}$    &$\Delta\bar{D}$      &$\Delta\bar{D}^{*}$  &~~~~~~~~~  \\
$I=2$    &$S=\frac{5}{2}$    &$\Delta\bar{D}^{*}$  &     &~~~~~~~~~\\
\bottomrule[1pt]
\end{tabular}
\end{center}
\end{table}

\subsection{The effective potentials}
\begin{figure*}
\includegraphics[scale=0.9]{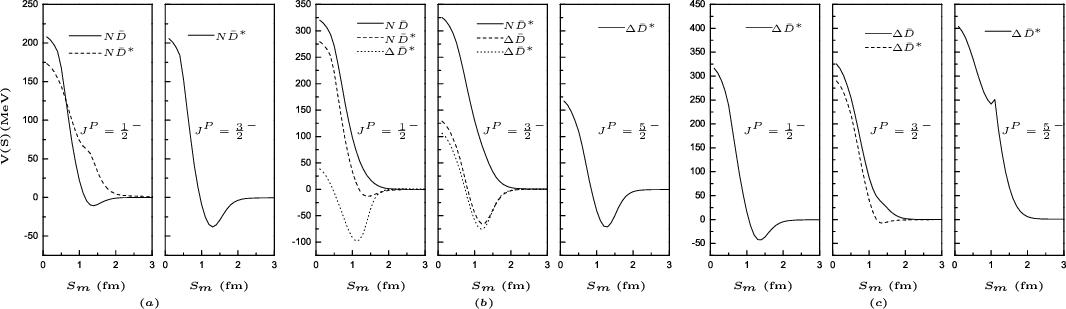}
\vspace{-0.5cm} \caption{The effective potentials of the $B\bar{D}$ system in QDCSM, and (a) $I=0$, (b) $I=1$, (c) $I=2$.  }
\label{Veff-t}
\end{figure*}

To search the possible pentaquark states, we estimate the effective potentials between these hadron pairs for the first step. Here the definition of potential can be written as
\begin{eqnarray}
E(S_{m})&=&\frac{\langle\Psi_{5q}(S_m)|H|\Psi_{5q}(S_m)\rangle}{\langle\Psi_{5q}(S_m)|\Psi_{5q}(S_m)\rangle},\label{Eq:PotentialE}
\end{eqnarray}
where $S_{m}$ stands for the distance between two clusters. $\Psi_{5q}(S_m)$ represents the wave function of a certain channel. Besides, $\langle\Psi_{5q}(S_m)|H|\Psi_{5q}(S_m)\rangle$ and $\langle\Psi_{5q}(S_m)|\Psi_{5q}(S_m)\rangle$ are the Hamiltonian matrix and the overlap of the states. So the effective potential between two colorless cluster is defined as,
\begin{eqnarray}
	V(S_m)=E(S_m)-E(\infty), \label{Eq:PotentialV}
\end{eqnarray}
where $E(\infty)$ stand for at a sufficient large distance of two clusters. The estimated effective potentials of the $B\bar{D}$ systems with $I=0$, $I=1$ and $I=2$ are shown in Fig.~\ref{Veff-t}.

For the $I(J^{P})=0(\frac{1}{2}^{-})$ system, there are two physical channels: $N\bar{D}$ and $N\bar{D}^{*}$. From the Fig.~\ref{Veff-t}(a), the effective potential of the $N\bar{D}$ shows attraction. So it is more possible for the $N\bar{D}$ channel to form a bound state although the interactions of $N\bar{D}$ is weakly. For the $N\bar{D}^{*}$ channels, the effective potentials of which is repulsive. For the $I(J^{P})=0(\frac{3}{2})^{-}$, one can see that the $N\bar{D}^{*}$ channel is attractive, so this channel may be a bound state, and a dynamic calculation about the $B\bar{D}$ system would be needed.

For the $I(J^{P})=1(\frac{1}{2})^{-}$ system, we can see that the effective potentials are attractive for the $N\bar{D}^{\ast}$ and $\Delta \bar{D}^{\ast}$ channels, while the one for $N\bar{D}$ channel is repulsive. It is obvious that the attraction of the $\Delta \bar{D}^{\ast}$ is the largest one, followed by the attraction of $N\bar{D}$ channels. So the $\Delta \bar{D}^{\ast}$ channel is most likely to form a bound state. For the $I(J^{P})=1(\frac{3}{2})^{-}$ system, From the Fig.~\ref{Veff-t}(b), the potentials for the $\Delta \bar{D}$ and $\Delta \bar{D}^{*}$ channels are attractive, while the potential for the $N\bar{D}^{\ast}$ channel is repulsive. So no bound state or resonance state can be formed in this channel. However, the bound states or resonance states are possible for the $\Delta \bar{D}$ and $\Delta \bar{D}^{*}$ channels due to the attractive nature of these interactions between the two clusters. For the $I(J^{P})=1(\frac{5}{2})^{-}$ system, only one channel $\Delta \bar{D}^{*}$ present, the interaction of $\Delta$ and  $\bar{D}^{*}$ shows a great attraction, which implies the possible bound state or resonance state in this channel.

 For the $I(J^{P})=2(\frac{1}{2})^{-}$ system and $I(J^{P})=2(\frac{5}{2})^{-}$ system, there is one channel $\Delta \bar{D}^{*}$, but the effective potentials of the $\Delta \bar{D}^{*}$ channel for this two system, is exactly opposite, and the effective potential of the former is attractive and the latter is repulsive. So it is interesting to explore the possibility of formation of bound or resonance state. For the $IJ^{P}=2\frac{3}{2}^{-}$ system, the effective potential of the $\Delta \bar{D}$ channel shows an repulsive property, while the $\Delta \bar{D}^{*}$ channel is attractive, and a dynamic calculation is needed here to check for existence of bound or resonance states.

\subsection{Possible bound states}
To check whether the possible bound state exist, a dynamic calculation is a very important process. It is important to note herein that just $S$-wave for the $B\bar{D}$ system would be taken into account. Moreover, a very efficient method to solve the bound state problem of a few body system is to employ the resonating group method(RGM)~\cite{Kamimura:1981oxj,Simons:1981gbz,Hiyama:2005cf,Hiyama:2018ukv}. For the RGM, firstly, the relative motion wave function between two clusters can be expanded by the Gaussian, then the integro-differential equation of RGM can be transformed into an algebraic equation, the generalized eigenequation. Finally, The intrinsic energy of the $B\bar{D}$ system is obtained by solving the Schrodinger equation. The details of RGM are shown in the Appendix~\ref{app}. In the calculation, two clusters separation $(|S_{m}|)$ is taken to be less than 6 fm (to keep the matrix dimension manageably small). For the $B\bar{D}$ system, all possible physical channels for each $I(J^{P})$ quantum numbers are listed in Table~\ref{channels}. Table~\ref{bound} summarize our calculated results of the lowest-lying $uudu\bar{c}$ pentaquark states. In this table, $E_{sc}$ and $E_{cc}$ are the eigenenergies of the $B\bar{D}$ system by the single channel estimations and the coupled channel estimations. $E_{th}^{Model}$ and $E_{th}^{Exp}$ stand for the theoretical estimations and experimental measurements of the thresholds of the channels. $E_{B}$ can be obtained by subtracting the difference between $E_{sc}$ and $E_{th}^{Model}$. Considering the uncertainties caused by the QDCSM estimations, the corrected eigenenergies of the single channel $E_{sc}^{\prime}$ can be obtained through the use of $E_{th}^{Exp}+E_{B}$. In a very similar way, we take the lowest the threshold of the involved channels as a scale, the corrected eigenenergy of the coupled channels $E_{cc}^{\prime}$ can be obtained.

For the $I(J^P)=0(\frac{1}{2}^-)$ system, the single-channel calculation shows that $N\bar{D}$, the attraction of which is too weak to tie the two particles, is unbound. For the $N\bar{D}^{\ast}$ channel, which also is an unbound state because of the repulsive nature of the interaction of $N$ and $\bar{D}^{\ast}$. However, when we consider the effect of  multichannel coupling, the lowest energy obtained is approximately $2802$ MeV, which is below the $N\bar{D}$ threshold with the binding energy of $-1$ MeV. So there is a bound state for the $I(J^P)=0(\frac{1}{2}^-)$ system. This results consistent with the results of the Refs.~\cite{Yasui:2009bz,Yamaguchi:2011xb}, in which the stability of $N\bar{D}$ in the $J^{P}=\frac{1}{2}^{-}$ with the $I=0$ was discussed.
 For the $I(J^P)=0(\frac{3}{2}^-)$ system, only one channel $N\bar{D}^{\ast}$ exists. Although the effective potential of the $N\bar{D}^{\ast}$ presents an attractive characteristic, the result of the bound calculation from the Table~\ref{bound} shows that the channel $N\bar{D}^{\ast}$ is not a bound state because their interactions are not sufficient to form a bound state.

For the $I(J^P)=1(\frac{1}{2}^-)$ system, there are three channels: $N\bar{D}$, $N\bar{D}^{*}$, and $\Delta\bar{D}^{\ast}$. The single channel bound calculation of the $N\bar{D}$ and $N\bar{D}^{\ast}$ indicate that these two channels are all unbound. As shown in Fig.~\ref{Veff-t}(b), the effective potential of the $N\bar{D}$ is repulsive, and the $N\bar{D}^{*}$ channel has only too weakly attraction to form a bound state. So it is reasonable. However, the $\Delta\bar{D}^{*}$ can make a bound state through the stronger attraction of the $\Delta$ and $\bar{D}^{\ast}$, and in comparison with the theoretical threshold of $\Delta$ and $\bar{D}^{\ast}$ channel, the lowest energy of this state is $3203$ MeV, with the binding energy of $-36$ MeV. After the multichannel coupling calculation, the lowest energy is still above the threshold of the $N\bar{D}$ channel, and it means that there is no bound state for the $I(J^{P})=1(\frac{1}{2}^{-})$ system. However, for the closed channel $\Delta\bar{D}^{\ast}$ channel, when considering the scattering process calculated with other open channels such as $N\bar{D}$, or $N\bar{D}^{\ast}$ , $\Delta\bar{D}^{\ast}$ may be a resonance state. So in the next section, we will study the scattering processes of these open channels to determine whether $\Delta\bar{D}^{\ast}$ can form a resonance state.

For the $IJ^P=1(\frac{3}{2}^-)$ system, which includes three channels: $N\bar{D}^{\ast}$, $\Delta\bar{D}$ and $\Delta\bar{D}^{\ast}$. The $\Delta\bar{D}$ and $\Delta\bar{D}^{\ast}$ are bound, and the lowest energy of those channels are below their theoretical thresholds, and from the Table~\ref{bound}, the binding energies are $-14$ MeV and $-26$ MeV, respectively. For the $N\bar{D}^{\ast}$ channel, no bound state can be found due to the repulsive effect of the interaction of the $N$ and $\bar{D}^{\ast}$.  When considering the three-channel coupling calculation of $N\bar{D}^{\ast}$, $\Delta\bar{D}$, $\Delta\bar{D}^{\ast}$, the result turns out that the lowest eigenvalue obtained is still higher the theoretical threshold of the corresponding physical channel, which means that there are no bound states. However, we should check if the $\Delta\bar{D}$ and $\Delta\bar{D}^{\ast}$ channel are resonance states by coupling the open channel $N\bar{D}^{\ast}$ in the next section. For the $I(J^P)=1(\frac{5}{2}^-)$ system, stable energy is obtained by the bound calculation, the mass of which is $-28$ MeV lower than the threshold of $\Delta\bar{D}^{\ast}$. So it is possible to a bound state. From the very strong attraction effect of $\Delta\bar{D}^{\ast}$, the result can be deemed valid. From the Ref.~\cite{Carames:2012bd}, the only one bound state in the $\Delta \bar{D}^{\ast}$ (T,S)=(1,5/2) system was obtained, this result is similar to our results. 

For the $I=2$ system, the $\Delta\bar{D}$ with $J^P=\frac{1}{2}^{-}$ is a bound state with the binding energy of $-10$ MeV while the $\Delta\bar{D}^{\ast}$ with $J^P=\frac{5}{2}^{-}$ is unbound as a rejection properties of the two particles. For the $J^P=\frac{3}{2}^{-}$, there is no bound state no matter the single calculation or the multichannel coupling calculation.


\begin{table*}[htb]
\begin{center}
\renewcommand{\arraystretch}{1.5}
\caption{\label{bound} The binding energies and the masses of every single channel and those of channel coupling for the pentaquarks with different quantum number. The values are provided in units of MeV. }
\begin{tabular}{p{1.6cm}<\centering p{1.6cm}<\centering p{1.6cm}<\centering p{1.6cm}<\centering p{1.6cm}<\centering p{1.6cm}<\centering p{1.6cm}<\centering p{1.6cm}<\centering p{2.0cm}<\centering p{1.6cm}<\centering  }
\toprule[1pt]
$I(J^{P})$  & Channel & $E_{sc}$ &$E_{th}^{Model}$ &$E_{B}$ &$E_{th}^{Exp}$ &$E_{sc}^{\prime}$ & $E_{cc}/E_B$ & $E_{cc}^\prime$ \\
 \midrule[1pt]
\multirow{2}{*}{$0(\frac{1}{2}^{-})$} &$N\bar{D}$               &2808 &2804 &$+4$ &2803 &2807  &\multirow{2}{*}{2803/-1}&\multirow{2}{*}{2802}\\
                                      &$N\bar{D}^{\ast}$        &2845 &2839 &$+6$ &2946 &2952     \\
                           \midrule[1pt]
\multirow{1}{*}{0$(\frac{3}{2}^{-})$} &$N\bar{D}^{\ast}$        &2843 &2839 &$+4$  &2946 &2950 &\multirow{1}{*}{2843/+4}&\multirow{1}{*}{2950} \\
                           \midrule[1pt]
\multirow{3}{*}{1$(\frac{1}{2}^{-})$} &$N\bar{D}$               &2809 &2804 &$+5$  &2803 &2808 &\multirow{3}{*}{2808/+4}& \multirow{3}{*}{2807}\\
                                      &$N\bar{D}^{\ast}$        &2843 &2839 &$+4$  &2946 &2950     \\
                                      &$\Delta\bar{D}^{\ast}$   &3096 &3132 &$-36$ &3239 &3203     \\
                           \midrule[1pt]
\multirow{3}{*}{$1(\frac{3}{2}^{-})$} &$N\bar{D}^{\ast}$        &2845 &2839 &$+6$  &2946 &2952 & \multirow{3}{*}{2844/+5}& \multirow{3}{*}{2951}\\
                                      &$\Delta\bar{D}$          &3083 &3097 &$-14$ &3096 &3082     \\
                                      &$\Delta\bar{D}^{\ast}$   &3106 &3132 &$-26$ &3239 &3213     \\
                           \midrule[1pt]
\multirow{1}{*}{1$(\frac{5}{2}^{-})$} &$\Delta\bar{D}^{\ast}$   &3104 &3132 &$-28$ &3239 &3211 &\multirow{1}{*}{3104/-28}& \multirow{1}{*}{3211} \\
                           \midrule[1pt]
\multirow{1}{*}{2$(\frac{1}{2}^{-})$} &$\Delta\bar{D}^{\ast}$          &3122 &3132 &$-10$ &3239 &3229 &\multirow{1}{*}{3122/-10}& \multirow{1}{*}{3229} \\
                           \midrule[1pt]
\multirow{2}{*}{2$(\frac{3}{2}^{-})$} &$\Delta\bar{D}$          &3102 &3097 &$+5$  &3096 &3101 &\multirow{2}{*}{3101/+4} & \multirow{2}{*}{3100} \\
                                      &$\Delta\bar{D}^{\ast}$   &3136 &3132 &$+4$  &3239 &3243    \\
                           \midrule[1pt]
\multirow{1}{*}{2$(\frac{5}{2}^{-})$} &$\Delta\bar{D}^{\ast}$   &3138 &3132 &$+6$  &3239 &3245 &\multirow{1}{*}{3138/+6}&  \multirow{1}{*}{3245} \\
\bottomrule[1pt]
\end{tabular}
\end{center}
\end{table*}

\subsection{Possible resonance states}
\begin{figure}
\includegraphics[scale=0.55]{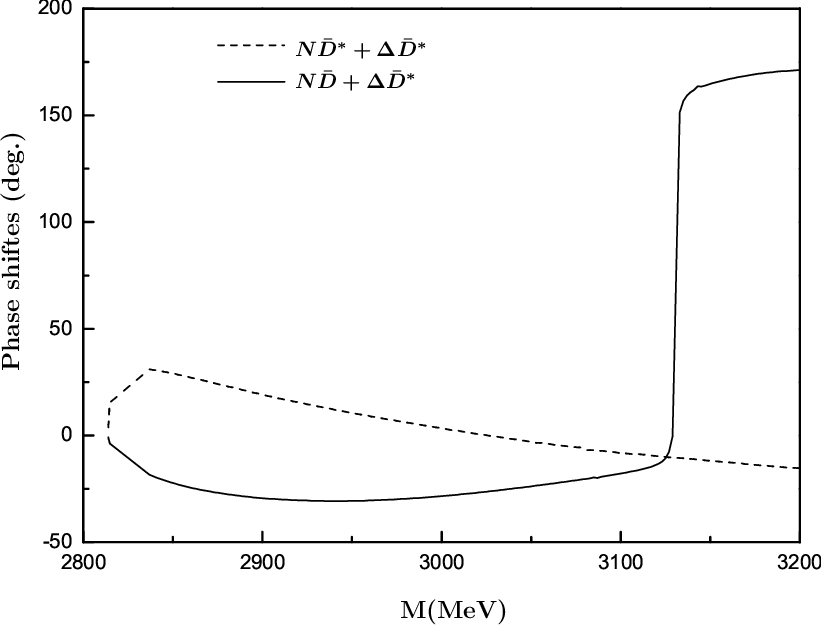}
\vspace{-0.2cm} \caption{The $N\bar{D}$ $S-$wave phase shifts with two-channel coupling for the $I(J^{P})=1(\frac{1}{2}^{-})$ system.}
\label{phase_11}
\end{figure}

\begin{figure}
\includegraphics[scale=0.55]{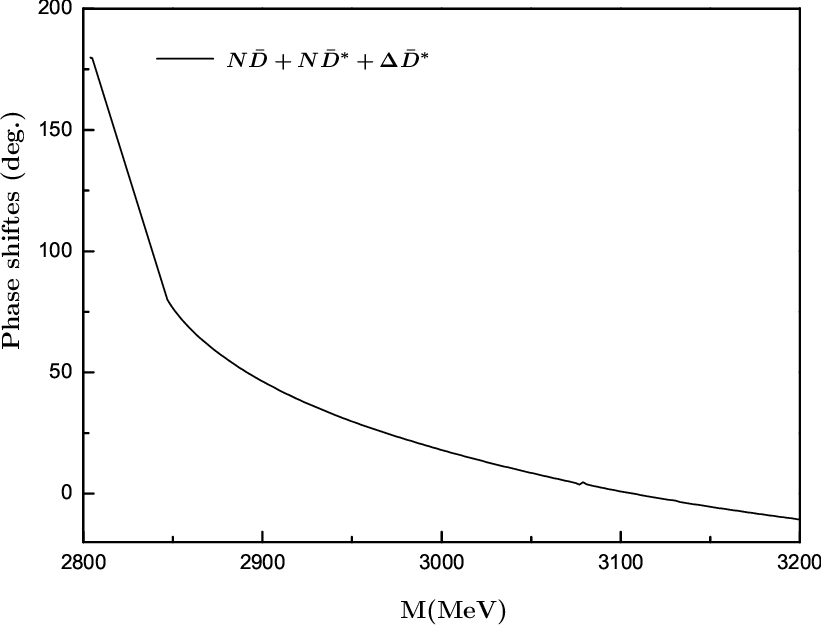}
\vspace{-0.2cm} \caption{The $N\bar{D}$ $S-$wave phase shifts with three-channel coupling for the $I(J^{P})=1(\frac{1}{2}^{-})$ system.}
\label{phase_11-T}
\end{figure}

\begin{figure}
\includegraphics[scale=0.55]{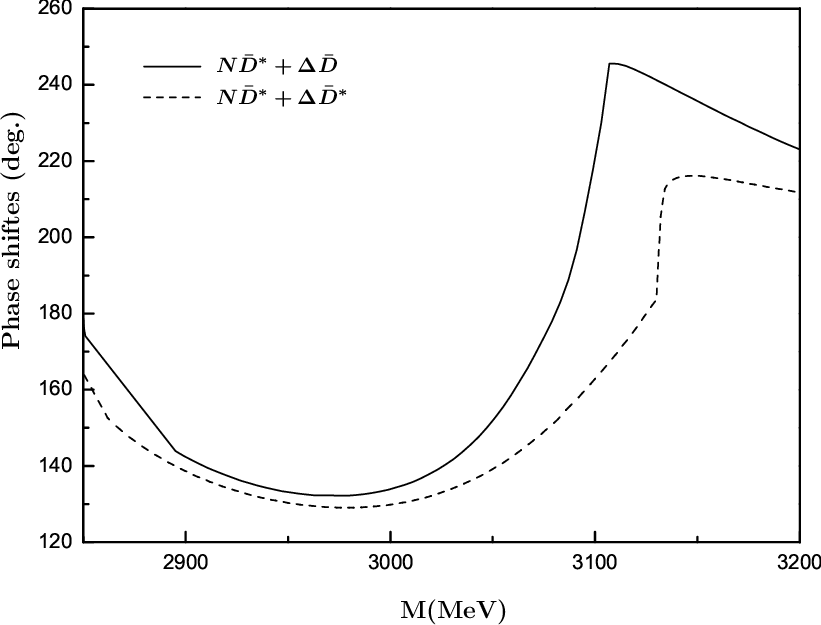}
\vspace{-0.2cm} \caption{The $N\bar{D}^{\ast}$ $S-$wave phase shifts with two-channel coupling for the $I(J^{P})=1(\frac{3}{2}^{-})$ system.}
\label{phase_13}
\end{figure}

\begin{figure}
\includegraphics[scale=0.55]{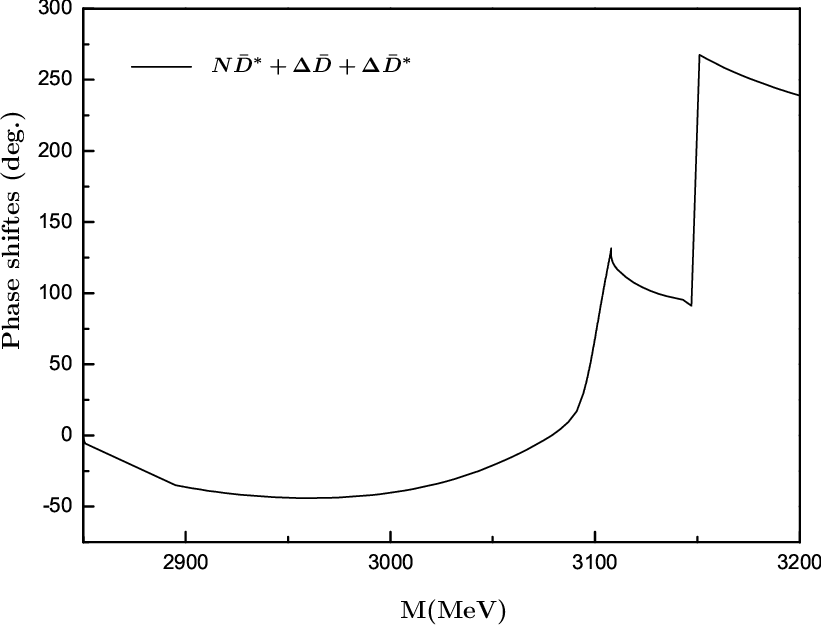}
\vspace{-0.2cm} \caption{The $N\bar{D}^{\ast}$ $S-$wave phase shifts with three-channel coupling for the $I(J^{P})=1(\frac{3}{2}^{-})$ system.}
\label{phase_13-T}
\end{figure}

In this section, to confirm whether or not these bound states can survive as resonance states after coupling to the open channels, the scattering phase shifts of all the open channels in QDCSM would be investigated. The details of the calculation method are shown in the Appendix A. Resonances are unstable particles usually observed as bell-shaped structures in scattering phase shifts process.  So the results of possible resonances are shown in Fig~\ref{phase_11} to Fig.~\ref{phase_13-T}. Here, we should note that the horizontal axis $M$ in Fig~\ref{phase_11} to Fig.~\ref{phase_13-T} is the sum of the corresponding theoretical threshold of the open channel and the incident energy. In addition, the resonance mass and decay width of the bound states in present the Table~\ref{decay1} where $M_{r}$ is the resonance mass, and $\Gamma_{i}$ is the partial decay width of resonance state, and $\Gamma_{total}$ is the total decay width. For the whole system, it is noted that all of the states that we study here are in the $S-$wave. Although the bound state of the $S-$wave can decay to a $D-$wave by tensor forces, its decay is almost negligible due to its small size, so the total decay widths of the states is smaller compared to the minimum limit. we proceed now to describe in detail our theoretical findings.

From the bound state calculation shown above, for the $I(J^{P})=1(\frac{1}{2}^{-})$ system, the $\Delta\bar{D}^{\ast}$ is bound while the $N\bar{D}$ and $N\bar{D}^{\ast}$ is unbound and those channels can be identified as the open channels. So the bound state $\Delta\bar{D}^{\ast}$ can decay to two open channels: $N\bar{D}$ and $N\bar{D}^{\ast}$. Firstly, we calculate the two-channel coupling. The phase shifts of all scattering channels are shown in Fig.~\ref{phase_11}. The phase shifts of the $N\bar{D}$ channel clearly show a resonance states, which means that the bound state $\Delta\bar{D}^{\ast}$ appears as a resonance state by coupling to the scattering channel $N\bar{D}$ while there is no indication of the presence of any resonance states in the scattering phase shift of the open channel $N\bar{D}^{\ast}$. Besides, we also consider the channel coupling calculation of $\Delta\bar{D}^{\ast}$ and two open channels $N\bar{D}$ and $N\bar{D}^{\ast}$. From Fig.~\ref{phase_11-T}, no resonance states are found, because the effect of the channel coupling raise the energies of $\Delta\bar{D}^{\ast}$ above its threshold. Moreover, in order to minimize the theoretical errors and to compare calculated results to the future experimental data, we shifts the mass of the resonance state to $M_{r}=M-E_{Model}^{th}+E_{Model}^{exp}$. Taking the resonance state $\Delta\bar{D}^{\ast}$ in the $N\bar{D}$ channel as an example, the resonance mass is about $3123$ MeV as shown in Fig.~\ref{phase_11}, the theoretical threshold is $E_{Model}^{th}=3132$ MeV, and the experimental threshold is $E_{Model}=3239$ MeV. So the finial resonance mass is $M_{r}=3123-3132+3239=3230$ MeV. The resonance mass and decay width are listed in Table.~\ref{decay1}.

For the $I(J^{P})=1(\frac{3}{2}^{-})$ system, two types of channel coupling are to be taken into account in the calculation. The first is the two-channel coupling with a single bound state and a related open channel, while the other is the three-channel coupling with two bound states and a corresponding open channel. Firstly, the two channel coupling is considered, and the phase shifts of all scattering channels are shown in Fig.~\ref{phase_13}. The phase shifts of the $N\bar{D}^{\ast}$ channel clearly show two resonance states, which means that the bound states $\Delta\bar{D}$ and $\Delta\bar{D}^{\ast}$ appear as a resonance state by coupling to the scattering channel $N\bar{D}^{\ast}$. The resonance mass and the decay width of every resonance state can be obtained from the shape of the resonance, which are listed in Table~\ref{decay1}, which shows that the $\Delta\bar{D}$ and $\Delta\bar{D}^{\ast}$  are all resonance state and the mass of those resonance are 3089 MeV and 3233 MeV with the decay width of 20 MeV and 4 MeV, respectively.  In addition, to investigate the effect of the coupling channel of the bound states, we also calculate the three channel coupling. The phase shifts of all scattering channels of the $I(J^{P})=1(\frac{3}{2}^{-})$ system are shown in Fig.~\ref{phase_13-T}, which shows a multi-resonance behavior. There are two resonance states in the $N\bar{D}^{\ast}$ scattering phase shifts corresponding to $\Delta\bar{D}$ and $\Delta\bar{D}^{\ast}$, respectively. The resonance masses and decay widths of resonance states with three channel coupling are also listed in Table~\ref{decay1}. The resonance mass of the $\Delta\bar{D}$ state is 3090 MeV and the decay width is about 14 MeV, and the $\Delta\bar{D}^*$ state has the mass of 3230 MeV and the decay width of 1 MeV.

\begin{table*}[!htb]
\renewcommand{\arraystretch}{1.2}
\caption{The resonance mass and decay width (in MeV) of the molecular pentaquarks with $I(J^{P})=1(\frac{1}{2}^{-})$ and $I(J^{P})=1(\frac{3}{2}^{-})$  in QDCSM.}\label{decay1}
\begin{tabular}{p{2.5cm}<\centering p{1.cm}<\centering p{1.cm}<\centering p{0.1cm}<\centering p{1.cm}<\centering p{1.cm}<\centering p{0.1cm}<\centering p{1.cm}<\centering p{1.cm}<\centering  p{0.1cm}<\centering  p{1.cm}<\centering p{1.cm}<\centering  p{0.1cm}<\centering  p{1.cm}<\centering  p{1.cm}<\centering }
\hline\hline
\multicolumn{1}{c}{} & \multicolumn{8}{c}{two-channel coupling} &\multicolumn{1}{c}{} & \multicolumn{5}{c}{three-channel coupling} \\
\cline{2-9}\cline{11-15}\\
\multicolumn{1}{c}{} &\multicolumn{2}{c}{$\Delta\bar{D}(J^{P}=\frac{3}{2}^{-})$} &\multicolumn{1}{c}{} &\multicolumn{2}{c}{$\Delta\bar{D}^{\ast}(J^{P}=\frac{3}{2}^{-})$} &\multicolumn{1}{c}{} &\multicolumn{2}{c}{$\Delta\bar{D}^{\ast}(J^{P}=\frac{1}{2}^{-})$} &\multicolumn{1}{c}{} &\multicolumn{2}{c}{$\Delta\bar{D}(J^{P}=\frac{3}{2}^{-})$} &\multicolumn{1}{c}{} &\multicolumn{2}{c}{$\Delta\bar{D}^{\ast}(J^{P}=\frac{3}{2}^{-})$}\\
\cline{2-3}\cline{5-6}\cline{8-9}\cline{11-12}\cline{14-15}\\
{open channels}  & $M_{r}$ &$\Gamma_{i}$  & &$M_{r}$ &$\Gamma_{i}$ &   &$M_{r}$ &$\Gamma_{i}$  & &$M_{r}$ &$\Gamma_{i}$ &  &$M_{r}$ & $\Gamma_{i}$ \\
\hline
 $N\bar{D}$     & $-$  & $-$     &      &$-$    &$-$   &      &3230  &0.01   &     &$-$   &$-$   &  &$-$   &$-$   \\
 {$N\bar{D}^{\ast}$}       & 3089  & 20     &      &3233   &4    &   &$-$    &$-$     &      &3090  &14    &  &3230  &1     \\
 {$\Gamma_{total}$}     &  ~~   & 20     &~~~   &~~~    &4    &      &~~~   &0.01   &      &~~~   &14    &  &~~~   &1     \\

\hline\hline
\end{tabular}
\end{table*}

\subsection{Low-energy scattering parameters }
\begin{figure*}[t]
    \centering
    \includegraphics[scale=0.25]{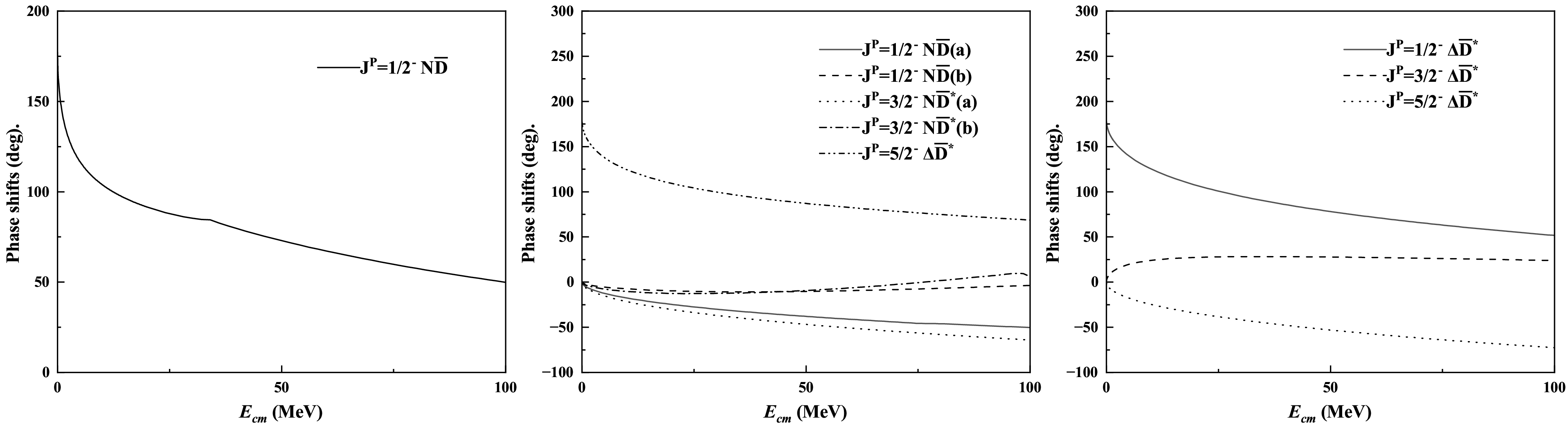}   
    \caption{The low-energy phase shifts of the open channel for different quantum numbers. Left panel: The low-energy phase shifts of $N\bar{D}$ obtained from two-channel coupling with $N\bar{D}$ and $N\bar{D}^{\ast}$. Middle panel: The labels (a) and (b) denote the single-channel and coupled-channel results, respectively. Specifically, $N\bar{D}(b)$ represents the  low-energy phase shifts of $N\bar{D}$ from two-channel coupling with  $N\bar{D}$ and $\Delta \bar{D}^{\ast}$; $N\bar{D}^{\ast}(b)$ shows the low-energy phase shifts of $N\bar{D}^{\ast}$ from the three-channel coupling with $N\bar{D}^{\ast}$, $\Delta \bar{D}$ and $\Delta \bar{D}^{\ast}$; for $J^{P}=5/2^{-}$, the low-energy phase shifts of $\Delta \bar{D}^{\ast}$ are presented. Right panel: The single-channel low-energy phase shifts results for $\Delta \bar{D}^{\ast}$ . }
    \label{low-phaseshifts-fig}
\end{figure*}

To bridge our QDCSM predictions with future measurements at ALICE and LHCb,  we investigate the two-particle momentum correlation functions for the  $N\bar{D}$ system in this section. However, before investigating the correlation functions, we first extract the low-energy scattering parameters, specifically the scattering length and effective range, from the calculated S-wave phase shifts using the effective range expansion:
\begin{eqnarray}\label{wave13}
    k \cot{\delta_L} &=& -\frac{1}{a_{0}}+\frac{1}{2}r_{0}k^{2}+O(k^4).
\end{eqnarray}
The extracted scattering parameters for various channels, along with the calculated binding energies ($E_{B}^{'}$), are summarized in Table~\ref{low-phaseshifts}. The corresponding low-energy phase shifts are displayed in Fig.~\ref{low-phaseshifts-fig}. 

For the $I(J^{P})=0(1/2^{-})$ system, the results demonstrate the crucial role of channel coupling . As shown in Table~\ref{low-phaseshifts}, the single -channel $N\bar{D}$ calculation yields a negative scattering length of $a_{0}=-1.17$ fm, indicating an attractive but unbound interaction. However, when considered the channel coupling with the $N\bar{D}^*$ channel, the scattering length shifts to a larger positive values of $a_{0}=3.88$ fm, and a bound state emerges with a binding energy of $E_{B}^{'}=-2.38$ MeV. This is consistent with the behavior of the phase shifts shown in the left panel of Fig.~\ref{low-phaseshifts-fig}, where the scattering phase shifts are positive and approach $180^\circ$ as the incident energy approaches zero, suggesting the existence of a bound state. 

In the $I=1$ sector,  the scenarios differ significantly across different total angular momenta. For the $J^{P}=1/2^-$ and  $J^P=3/2^-$, the scattering lengths are relatively small and positive, indicating that the effective interaction at the very threshold is dominated by a repulsive background. However, the phase shifts in the middle panel of Fig.~\ref{low-phaseshifts-fig} reveal a more complex dynamical structure induced by channel coupling. While these phase shifts start at  $0^\circ$ and initially exhibit negative values consistent with the repulsive background, they show an upward trend (transitioning from negative to positive) as the energy increases. This behavior serves as a clear signature of the resonance discussed in Section.C. These resonances originate from the deeply bound $\Delta \bar{D}$ and $\Delta \bar{D}^{\ast}$ states, which appears as resonances in the open channels $N \bar{D}$ and $N \bar{D}^{\ast}$ scattering continuum at higher energies. In contrast, for the $J^{P}=5/2^{-}$ case,  the interaction in the $\Delta \bar{D}^{\ast}$  channel is strongly attractive. This attraction is sufficient to support a deep bound state with a binding energy of $E^{'}_{B}=-22.2$ MeV. The corresponding scattering length is $a_{0}=1.86$ fm. It is worth noting that while still positive, this value is smaller than of the $I=0$ shallow bound state, reflecting the more compact nature of this deeply bound system.

For the $I=2$ sector, the results highlight a strong spin dependence of the interaction. For the $J^{P}=1/2^-$, the phase shifts of $\Delta\bar{D}^{\ast}$ channel starts at $180^\circ$ at zero energy and decreases as energy increases. This behavior is a definitive signature of the existence of a single bound state, and this corroborated by the extracted positive scattering length, which reflects effective size of bound system. For the $J^{P}=3/2^-$ and $J^{P}=5/2^-$, $\Delta\bar{D}^{\ast}$ channel is unbound with phase shifts starting at $0^\circ$. Specifically, the $J^{P}=3/2^-$ channel exhibits a negative scattering length with positive phase shifts, indicating a weak attractive interaction, while the $J^{P}=5/2^-$ shows a positive scattering length with negative phase shifts, corresponding to a repulsive interaction.

It is instructive to compare our derived scattering parameters for the $N\bar{D}$ system with other theoretical approaches summarized in Ref.~\cite{Barbat:2025orm}. As shown in that work, most effective field theory models based on Weinberg-Tomozawa or Zero-Range approximations predict a near-zero or repulsive scattering length for the $I=0$ $N\bar{D}$ channel. However, our QDCSM result aligns with the class of models that predict a strong attraction or a bound state, such as the Yamaguchi model. The large magnitude of the scattering length is a common feature when a bound state is located close to the threshold, suggesting that the $N\bar{D}$ interaction might be stronger than what is predicted by simple leading-order chiral effective theories. In addition to the bound state sector, checking the consistency in the open channels is equally important. For the $I=1$ $N\bar{D}$ channel, theoretical models listed in Ref.~\cite{Barbat:2025orm} universally predict a repulsive interaction. Our calculation yields a positive scattering length for the coupled-channel case. Although the sign convention for repulsion differs between our work  and the models in Ref.~\cite{Barbat:2025orm}, the physical conclusion is identical: the $I=1$ interaction at the threshold is repulsive and relatively weak. This agreement suggests that our model correctly captures the repulsive core in the non-resonant isospin sector, lending further credence to its prediction of attraction in the $I=0$ sector.

These extracted scattering parameters quantify the interaction strength in each channel and serve as the basis for constructing the equivalent potentials used to calculate the femtoscopic correlation functions in the following section.

\begin{table*}[htb]
    \begin{center}
        \renewcommand{\arraystretch}{1.5}
        \caption{\label{low-phaseshifts} The scattering length $a_{0}$, the effective range $r_{0}$, the binding energy $E_{B}^{\prime}$ determined by the variation method and the potential parameters used in the two-range Gaussian parametrization. Noted that the "a" in parentheses indicates consideration of a single channel only, while the "b" denotes the lowest open channel after channel coupling. Other unmarked case also represent single-channel scenarios. }
        \begin{tabular}{C{1.1cm} C{1.2cm}  C{1.5cm} C{1.5cm} C{1.5cm} C{1.5cm} C{1.5cm} C{1.5cm} C{1.5cm}}
            \toprule[1pt]
            $I(J^{P})$ &    channel&    $a_{0}$ (fm)&     $r_{0}$ (fm)&     $E_{B}^{\prime}$ (MeV)  & $V_1$ (MeV) & $\mu_1$ (fm) & $V_2$ (MeV) & $\mu_2$ (fm) \\
            \midrule[1pt]
            $0(1/2^{-})$&    $N \bar{D}(a)$                         &-1.17      &1.51     & $\ldots$      &- 589.63     &0.53     &947.37       &0.38\\
            $0(1/2^{-})$&    $N \bar{D}(b)$                         &3.88      &0.48     & -2.38            &-1158.34   &0.85   &3913.58   &0.23\\\\
            $1(1/2^{-})$&    $N \bar{D}(a)$                         &0.54      &0.45     &  $\ldots$      &-1903.13    &0.65   &4892.52   &0.28\\
            $1(1/2^{-})$&    $N \bar{D}(b)$                         &0.25      &-1.76    &  $\ldots$     &-838.96      &0.24   &88.49   &0.74\\
            $1(3/2^{-})$&    $N \bar{D}^{\ast}(a)$              &0.66      &0.49     &  $\ldots$     & -3031.84  &0.48   &1234.14   &0.62\\
            $1(3/2^{-})$&    $N \bar{D}^{\ast}(b)$              &0.35      &-0.97    &  $\ldots$    &-3571.23   &0.20   &505.23   &0.58\\
            $1(5/2^{-})$&    $\Delta \bar{D}^{\ast}$           &1.86      &0.56     &   -22.2        &-3125.45   &0.46   &299.81   &0.81\\
            $2(1/2^{-})$&    $\Delta \bar{D}^{\ast}$           &1.69      &0.77     &  -11.9         &-539.68   & 0.65  &473.05   &0.47\\
            $2(3/2^{-})$&    $\Delta \bar{D}^{\ast}$          &-0.92     &1.59     &  $\ldots$    &-200.56   &0.57   &2966.97   &0.17\\
            $2(5/2^{-})$&    $\Delta \bar{D}^{\ast}$          &0.70      &0.63     &  $\ldots$     &-758.09   &0.75   &8032.11   &0.55\\
            \bottomrule[1pt]  
        \end{tabular}
    \end{center}
\end{table*}

\subsection{ Femtoscopic correlation functions}

Based on the above S-wave scattering phase shifts extracted, we construct effective local potentials that reproduce the low-energy phase shifts obtained above. We employ a double-Gaussian form for the effective potential,  $V_{eff}(r)=V_{1}e^{-r^2/\mu_{1}^{2}}+V_{2}e^{-r^2/\mu_{2}^{2}}$, which is flexible enough to capture the short-range dynamics.  The fitted potential parameters for the representative channels are listed in Table~\ref{low-phaseshifts}. Subsequently, these potentials are incorporated into the CATS framework~\cite{Mihaylov:2018rva}. By numerically solving the Schr$\ddot{o}$dinger equation with these potentials, we obtain the the relative wave function of the baryon-meson pair, which is then integrated with source function to compute the theoretical correlation functions $C(K)$.

\begin{figure*}[t]
    \centering
    \includegraphics[scale=0.25]{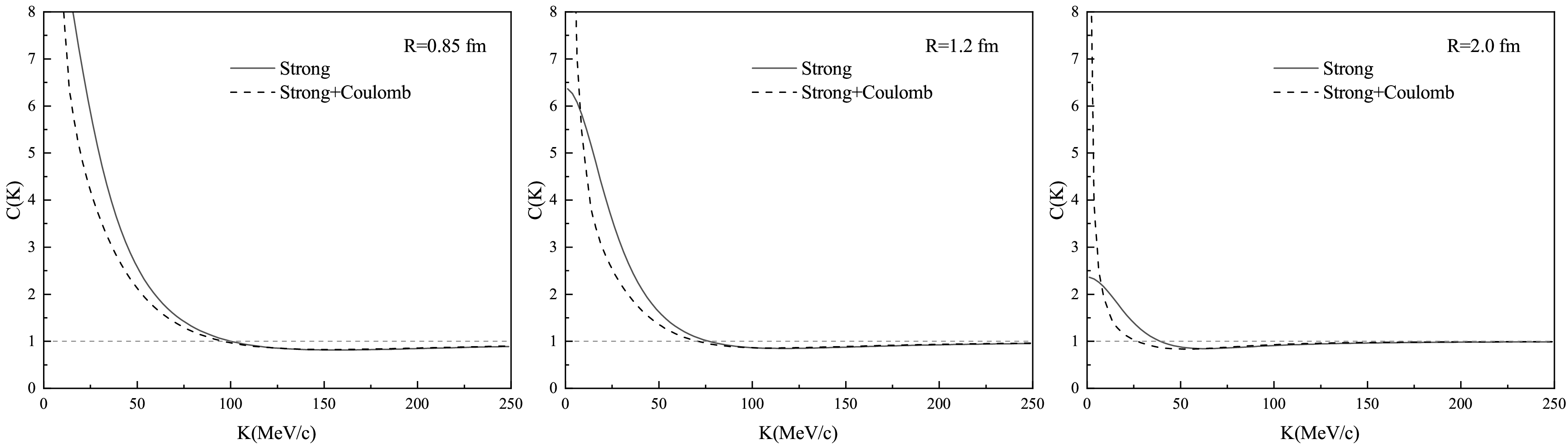}   
    \caption{Correlation functions of $p\bar{D}^{-}$  for different emitting source sizes in $I=0,I_{z}=0$ case.}
    \label{I=0-CF}
\end{figure*}

To investigate the femtoscopic observables for the $I(J^{P})=0(1/2^{-})$ state, where a bound state is predicted, we consider two physical scenarios: pure strong interaction and full interaction including the attractive coulomb force. It is worth noting that the full interaction scenario actually studies the physical $p D^{-}$ particle pair with $I=0$ based on dynamical estimation, although the $p D^{-}$ system involves a mixture of $I=0$ and $I=1$ components. The results are presented in Fig.~\ref{I=0-CF} for three different source size. As shown in Fig.~\ref{I=0-CF}, a common feature obtained in both scenarios is the significant enhancement of the correlation function in the low-momentum region. However, as the momentum increases, the correlation curves rapidly drop below the baseline ($C(K)=1$), which serves as a characteristic signature for the existence of a bound state. This characteristic suppression is consistent with the recent theoretical analysis in Ref.~\cite{Barbat:2025orm} for the $p D^{-}$ system. In their work, the interaction model featuring a bound state (specifically the Yamaguchi model) yields a correlation function that is significantly suppressed compared to those derived from repulsive or weakly attractive effective field theories (see Fig. 2 in Ref.~\cite{Barbat:2025orm}). This consistency corroborates our conclusion that the $N\bar{D}$ bound state exists. Furthermore, Fig.~\ref{I=0-CF} exhibits a significant dependence on the source size: the correlation signals are strong for small source size but gradually weakens as the source size increases to 2 fm. Finally, a comparison between the two scenarios reveals that the inclusion of the coulomb interaction further enhances the correlation singles in the low-momentum region.

\begin{figure*}[t]
    \centering
    \includegraphics[scale=0.25]{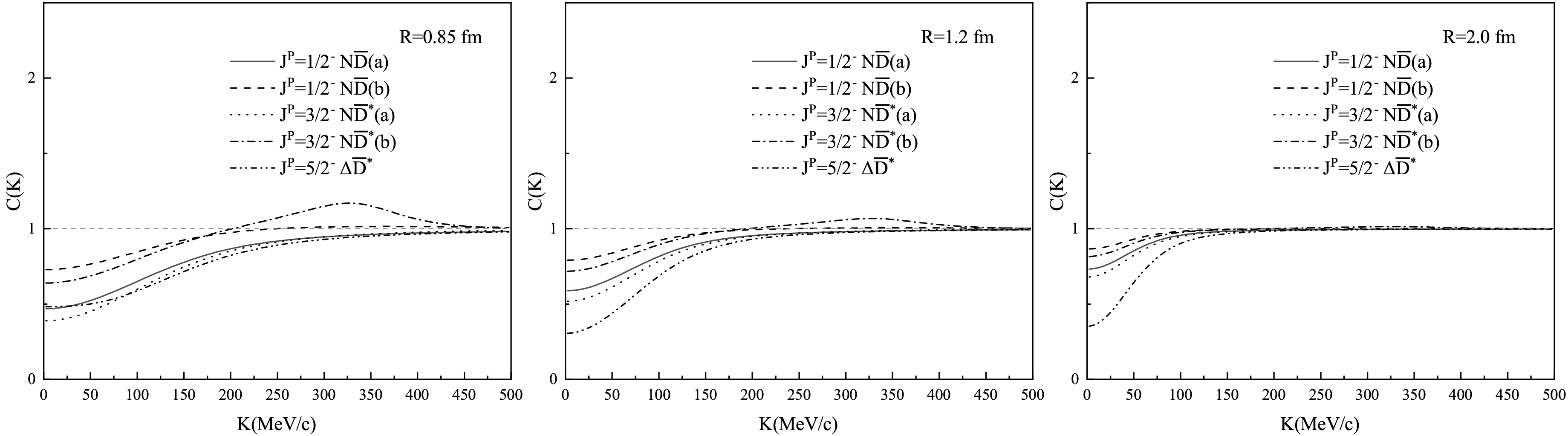}   
 \caption{Correlation functions of different channel  for different emitting source sizes in $I=1,I_{z}=1$ case. }
    \label{I=1-CF}
\end{figure*}

For the $I=1, I_{z}=1$ case, we systematically calculate the two-particle correlation functions for the $N\bar{D}$, $N \bar{D}^{\ast}$ and $\Delta \bar{D}^{\ast}$ channels.  Fig.~\ref{I=1-CF} illustrates the evolution of the correlation functions as a function of the relative momentum for three different emitting source size.  It is observed that in the low-momentum region, the correlation function of all considered channels are significantly below the baseline. This qualitatively reflects the dominance of the short-range hard-core repulsion in this isospin channel. Since $I_{z}=1$ corresponds to a pure isospin state and coulomb interaction are excluded in these calculations, this low-energy behavior originates purely from the dynamics of the strong interaction. For the $J^{P}=1/2^{-}$ and $J^{P}=3/2^{-}$ channels, the coupled curves (b) exhibit a distinct elevation relative to the single-channel curves (b), indicating that  the attractive effect generated by the channel coupling offsets the original repulsive interaction within a specific momentum region.  More importantly, at an emission source size R=0.85 fm, the channel coupling correlation function displays a distinct resonance peak structure, with values rising from below unity to above unity. This phenomenon reveals a significant enhancement of the wave function by the resonance state generated from channel coupling. It makes a transition in the nature of the interaction from repulsive to attractive, thereby providing direct dynamical evidence for the existence of this resonance. Furthermore, as the source size R increases, the resonance signal is gradually suppressed, and the correlation functions tend to flatten out. In the case of  $J^{P}=5/2^{-}$, the correlation functions exhibit a non-monotonic dependence on the source size, initially decreasing and then increasing. This specific behavior suggests a complex spatial overlap between the bound state wave function and the repulsive interaction, indicating that the system is highly sensitive to the size of the emission source.

\begin{figure*}[t]
    \centering
    \includegraphics[scale=0.25]{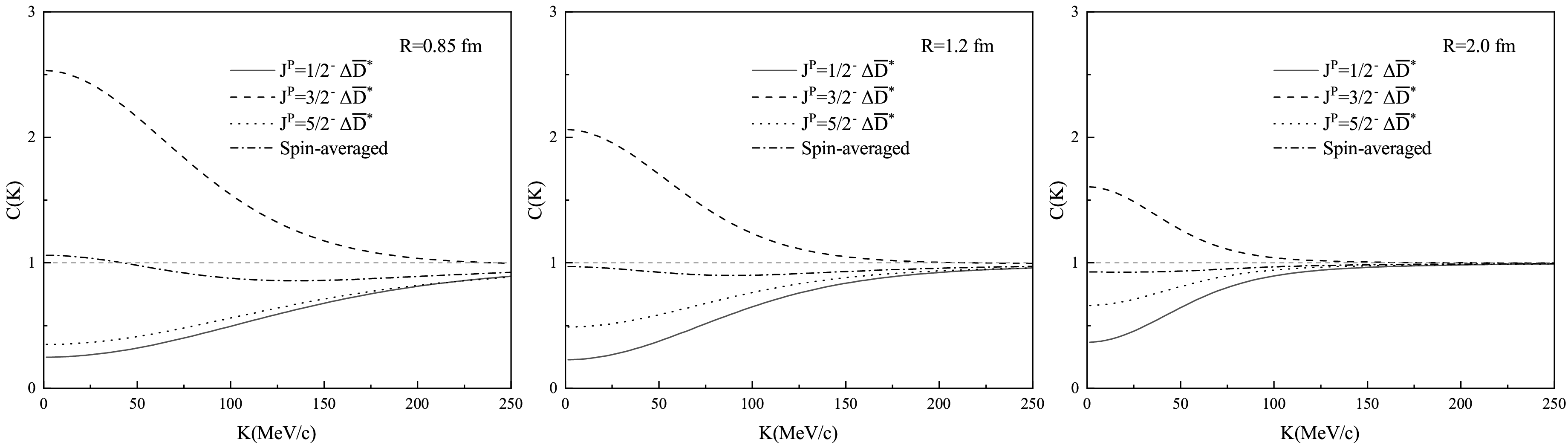}   
    \caption{Correlation functions of $\Delta\bar{D}^{\ast}$ for different emitting source sizes in $I=2,I_{z}=2$ case  }
    \label{I=2-CF}
\end{figure*}

For the $I=2, I_{z}=2$ case,  we investigate the $\Delta \bar{D}^{\ast}$ system, which is free from coulomb interactions, serving as a clean probe for strong forces.  As shown in Fig.~\ref{I=2-CF}, the femtoscopic observables reveal a pronounced sensitivity to the total angular momentum $J$, with distinct bahaviors evolving as the emitting source size increases from R=0.85 to 2.0 fm. The $J^{P}=1/2^{-}$ channel, which has a bound state, exhibits significant suppression across the low-momentum region. This characteristic suppression aligns with the femtoscopic behavior of deeply bound states reported in Ref.~\cite{Liu:2023uly}. Conversely, the unbound $J^{P}=3/2^{-}$ channel displays a clear enhancement above unity, consistent with an attractive interaction. Notably, this enhancement is sensitive to the source size; as R increases, the signal is progressively diluted, with the peak value decreasing significantly. For this specific configuration,  the single channel results are presented, as the dynamic analysis determined that coupled-channel effect exert negligible influence on the outcome. Meanwhile, the $J^{P}=5/2^{-}$ channel is suppressed, reflecting the repulsive nature of this high-spin configuration. Finally, the spin-averaged correlation, weighted by the statistical factor $(2J+1)$, demonstrates a cancellation effect: the enhancement from the attractive $J^{P}=3/2^{-}$ channel is largely offset by the suppression from $J^{P}=1/2^{-}$ and $J^{P}=5/2^{-}$ channels, yielding a result close to unity. This suggests that disentangling the rich microscopic dynamical of the $\Delta \bar{D}^{\ast}$ system in experiments requires spin-selective measurements.

\section{Summary\label{sum}}

  In the present work, we have systematically studied the $B\bar{D}$ system composed of $uudu\bar{c}$ in the framework of the quark delocalization color screening model (QDCSM). By solving the RGM equation, we attempt to explore if there are any possible bound and resonance states. Herein, the computed results indicate that several possible bound and resonance states are found for the $B\bar{D}$ system within the scanned quantum numbers: $I(J^{P})=0(\frac{1}{2}^{-})$, $I(J^{P})=1(\frac{1}{2}^{-})$, $I(J^{P})=1(\frac{3}{2}^{-})$, $I(J^{P})=1(\frac{5}{2}^{-})$, and $I(J^{P})=2(\frac{1}{2}^{-})$. These are characterized by the following features:

\begin{enumerate}[label=(\arabic*)]
    \item There are some bound states of $\Delta\bar{D}^{\ast}$ with $I(J^{P})=1(\frac{1}{2}^{-})$, $\Delta\bar{D}$ with $I(J^{P})=1(\frac{3}{2}^{-})$ and $\Delta\bar{D}^{\ast}$ with $I(J^{P})=1(\frac{3}{2}^{-})$ when considering the single-channel calculation. But when the coupling channel calculations are applied to the entire $B\bar{D}$ system , the $N\bar{D}$ for the $I(J^{P})=0(\frac{1}{2}^{-})$ can be turned into a bound state and the  $\Delta\bar{D}^{\ast}$ in the $I(J^{P})=0(\frac{5}{2}^{-}$) and $I(J^{P})=2(\frac{1}{2}^{-})$ are also supposed to the bound states.
    The presence of those states is also a sharp prediction of quark exchange dynamics because in a hadronic model the attraction appears in different channels.
    
    \item Narrow baryon-meson resonance states are obtained in coupled-channels cases, which are $\Delta\bar{D}^{\ast}$ with $I(J^{P})=1(\frac{1}{2}^{-})$ ($M_{r}=3220$ MeV, $\Gamma$=0.01 MeV) and $I(J^{P})=1(\frac{3}{2}^{-})$ ($M_{r}= 3230\sim3233$ MeV, $\Gamma$=$1\sim4 $MeV), and $\Delta\bar{D}$ with $I(J^{P})=1(\frac{3}{2}^{-})$ ($M_{r}= 3089\sim3090$ MeV, $\Gamma$=$14\sim20$ MeV). last but not least, we hope our sophisticated calculations of the pentaquarks may provide valuable information for the future experimental searches.
    
    \item Based on the derived scattering parameters, we successfully constructed the femtoscopic correlation functions. For the $I=0$ $N\bar{D}$ system, a distinct signal is predicted, characterized by a correlation strength significantly exceeding the baseline at low momenta and a suppression below the baseline at intermediate momenta. For the $I=1$ sector, the correlation functions reflect the dynamical influence of the resonance states generated by channel coupling, manifesting as distinct deviations from the single-channel repulsive background. Finally, for the $I=2$ sector, the results demonstrate a pronounced spin dependence, where the suppression signature of the $J=1/2^{-}$ bound state stands in contrast to the behaviors of other spin configurations. Therefore, we suggest that future experimental investigations at  ALICE should aim for spin-selective measurements to verify the existence of these predicted pentaquark states, as standard spin-averaged analyses might fail to detect them. 
\end{enumerate}

Beyond these predictions, it is worth highlighting the connection to recent experimental findings. Comparisons with ALICE data~\cite{ALICE:2022enj} analyzed in Ref.~\cite{Barbat:2025orm} suggest that the measured $D^-p$ correlation is best described by theoretical models featuring strong attraction or bound states, whereas models with weak interactions fail to reproduce the experimental trends. Our QDCSM calculation, which predicts an $I=0$ bound state and a consequent strong enhancement in the correlation function, aligns with this phenomenological preference. Therefore, the bound state predicted in this work represents a physically plausible scenario supported by the current experimental landscape, rather than a mere theoretical possibility.

For future investigations, we plan to refine the current model by incorporating tensor force contributions. This addition is crucial for a more precise determination of decay widths, especially for resonances involving $D$-wave. Moreover, the framework developed here—bridging the gap between quark-level dynamics and femtoscopic observables—can be readily extended to other heavy-flavor exotic systems, paving the way for a deeper understanding of hadron-hadron interactions in the era of high-precision experiments.

\acknowledgments{This work is supported partly by the National Natural Science Foundation of China under
    Contract Nos.12575088, 12175037, 12335001, 11775118, 11535005 and is also supported by the Natural Science Foundation for Youths of Henan Province  No. 252300421781.  School-Level Research Projects of Henan Normal university (No. 20240304) also supported this work.}

\appendix
\section{resonating group method for bound-state and scattering problems\label{app}}
In the present work, We perform bound state calculations as well as scattering calculations for the $N\bar{D_{s}}$ system by using the RGM in QDCSM. The issue of this approach is how to deal with the two-body problem. In this method, when dealing with the two-cluster system, one can only consider the relative motion between the clusters, while the two clusters are frozen inside. So the wave function of the baryon-meson system is

\begin{equation}
\label{wave1}
    \begin{split}
        \psi = \sum_{L}\mathcal{A} \Big[ & \left[\hat{\phi}_{A}\left(\boldsymbol{\rho_{A}},\boldsymbol{\lambda_{A}}\right)\hat{\phi}_{B}\left(\boldsymbol{\rho_{B}}\right)\right]^{[\sigma]IS} \\
        & \otimes \chi_{L}(\textbf{R}_{AB}) \Big]^{J},
    \end{split}
\end{equation}
where $L$ stands for the orbital angular momentum and the symbol $\mathcal{A}$ is the antisymmetry operator, which can be defined as
\begin{eqnarray}\label{wave2}
    \mathcal{A}&=& 1-P_{14}-P_{24}-P_{34},
\end{eqnarray}
where 1, 2, and 3 stand for the quarks in the baryon cluster, and 4 stands for the quark in the meson cluster. $\hat{\phi_{A}}$ and $\hat{\phi_{B}}$ are the internal cluster wave functions of the baryon A and meson B:

\begin{align}\label{wave3}
    \begin{split} 
        \hat{\phi}_{A} &= \left(\frac{2}{3\pi b^{2}}\right)^{3/4} \left(\frac{1}{2\pi b^{2}}\right)^{3/4} \\
        &\quad \times e^{-\left(\frac{\boldsymbol{\rho_{A}}^2}{4b^2}+\frac{\boldsymbol{\lambda_{A}}^2}{3b^2}\right)} \eta_{I_{A}}S_{A}\xi_{A}^{c},
    \end{split}\\
    \label{eq:A4}
    \hat{\phi}_{B} &= \left(\frac{1}{2\pi b^{2}}\right)^{3/4} e^{-\frac{\boldsymbol{\rho_{B}}}{4b^2}} \eta_{I_{B}}S_{B}\xi_{B}^{c},
\end{align}

where $\eta_{I}$, $S$, and $\xi$ represent the flavor, spin, and internal color terms of the cluster wave functions, respectively. $\rho_{A}$ and $\lambda_{A}$ are the internal coordinates for the baryon cluster A and $\rho_{B}$ is the internal coordinate for the meson cluster B. The Jacobi coordinates are defined as follows:
\begin{eqnarray}\label{wave4}
\boldsymbol{\rho_{A}}&=&\boldsymbol{r_{1}-r_{2}}, \ \ \ \boldsymbol{\rho_{B}}=\boldsymbol{r_{4}-r_{5}},\nonumber \\
\boldsymbol{\lambda_{A}}&=& \boldsymbol{r_{3}}-\frac{1}{2}(\boldsymbol{r}_{1}+\boldsymbol{r}_{2}),\nonumber \\
\boldsymbol{R_{A}}&=& \frac{1}{3}(\boldsymbol{r_{1}}+\boldsymbol{r_{2}}+\boldsymbol{r_{3}}), \ \ \ \boldsymbol{R_{B}}=\frac{1}{2}(\boldsymbol{r_{4}}+r_{5}),\nonumber \\
\boldsymbol{R_{AB}}&=&\boldsymbol{R_{A}}-\boldsymbol{R_{B}},\ \ \ \boldsymbol{R_{G}}=\frac{3}{5}\boldsymbol{R_{A}}+\frac{2}{5}\boldsymbol{R_{B}}.
\end{eqnarray}
From the variational principle, after variation with respect to the relative motion wave function $\chi\boldsymbol(R)=\sum_{L}\chi_{L}\boldsymbol(R)$, one obtains the RGM equation,
\begin{eqnarray}\label{wave5}
  \int H\left(\boldsymbol{R, R^{\prime}}\right)\chi\left(\boldsymbol{R^{\prime}}\right)d\boldsymbol\left(R^{\prime}\right)=E \nonumber \\  \times\int N\left(\boldsymbol{R, R^{\prime}}\right) \chi\left(\boldsymbol{R^{\prime}}\right)d\boldsymbol\left(R^{\prime}\right),
\end{eqnarray}
where $H(\boldsymbol{R, R^{\prime}})$ and $N(\boldsymbol{R, R^{\prime}})$ are Hamiltonian and norm kernels, respectively. The eigenenergy $E$ and the wave functions are obtained by solving the RGM equation.

Generally, one can introduce generator coordinates $S_{m}$ to expand the $L$th relative motion wave function $\chi_{L}(\boldsymbol{R})$ by\footnote{In the present estimation, only $S$-wave bound state is considered, i.e., $L=0$.},
 \begin{eqnarray}\label{wave6}
\chi_{L}\boldsymbol{(R)}&=&\frac{1}{\sqrt{4\pi}}\left(\frac{6}{5\pi b^2}\right)^{3/4}\sum_{m=1}^{n}C_{m}\nonumber\\
&&\int \exp\left[-\frac{3}{5b^2}\left(\boldsymbol{R}-\boldsymbol{S}_{m}\right)^2\right]Y^{L}\left(\hat{\boldsymbol{S}}_{m}\right)d\hat{\boldsymbol{S}}_{m} \nonumber\\
&=&\sum_{m=1}^{n}C_{m} \frac{u_{L}\left(\boldsymbol{R},\boldsymbol{S}_{m}\right)}{\boldsymbol{R}}Y^L\left(\hat{\boldsymbol{R}}\right),
\end{eqnarray}
with
\begin{equation}\label{wave7}
    \begin{split}
        u_{L}(\boldsymbol{R},\boldsymbol{S}_{m}) &= \sqrt{4\pi}\left(\frac{6}{5\pi b^2}\right)^{3/4} \boldsymbol{R} \, e^{-\frac{3}{5b^2}\left(\boldsymbol{R}-\boldsymbol{S}_{m}\right)^2} \\
        &\quad \times m^L j_{L}\left(-i\frac{6}{5b^2}S_{m}\right),
    \end{split}
\end{equation}
where $C_{m}$ is expansion coefficients, $n$ is the number of the Gaussian bases, which is determined by the stability of the results, and $j_{L}$ is the $L$th spherical Bessel function. Then the relative motion wave function $\chi(\textbf{R})$ is
\begin{align}\label{wave8}
    \chi(\textbf{R}) &= \frac{1}{\sqrt{4\pi}}\sum_{L}\left(\frac{6}{5\pi b^2}\right)^{3/4} \nonumber \\
    &\quad \times \sum_{m=1}^{n}C_{m}\int e^{-\frac{3}{5b^2}\left(\textbf{R}-\textbf{S}_{m}\right)^2}Y^L\left(\hat{\boldsymbol{S}}_{m}\right) d\hat{\boldsymbol{S}}_{m}.
\end{align}
After the inclusion of the center of mass motion,
\begin{eqnarray}\label{wave9}
\Phi_{G}(\textbf{R}_{G})&=&\left(\frac{5}{\pi b^2}\right)^{3/4}e^{-\frac{5}{2b^2}\textbf{R}_{G}},
 \end{eqnarray}
the total wave function in Eq.(\ref{wave1}) can be rewritten as,
\begin{equation}\label{wave10}
    \begin{split}
        \Psi_{5q} &= \mathcal{A}\sum_{m,L}C_{m,L}\int\frac{1}{\sqrt{4\pi}} \prod_{\alpha=1}^{3}\Phi_{\alpha}(S_{m})\prod_{\beta=4}^{5}\Phi_{\beta}(-S_{m}) \\
        &\quad \times \left[\left [\eta_{I_{A}S_{A}}\eta_{I_{B}S_{B}}\right]^{IS}Y^{L}(\hat{\boldsymbol{S}}_{m})\right]^J \left[\xi_{c}(A)\xi_{c}(B)\right]^{[\sigma]}.
    \end{split}
\end{equation}
where $\Phi_{\alpha}(S_{m})$ and $\Phi_{\beta}(-S_{m})$ are the single-particle orbital wave function with different reference centers, which specific form can be seen in Eq.~(\ref{wave0}).

With the reformulated ansatz as shown in Eq.~(\ref{wave10}), the RGM equation becomes an algebraic eigenvalue equation,
\begin{eqnarray}
  \sum_{j,L}C_{J,L}H_{i,j}^{L,L^{'}} &=& E\sum_{j}C_{j,L^{'}}N_{i,j}^{L^{'}},
\end{eqnarray}
where $N_{i,j}^{L^{'}}$ and $H_{i,j}^{L,L^{'}}$ are the  overlap of the wave functions and the matrix elements of the Hamiltonian, respectively. By solving the generalized eigenvalue problem, we can obtain the energies of the pentaquark $E$ and the corresponding expansion coefficient $C_{j,L}$. Finally, the relative motion wave function between two clusters can be obtained by substituting the $C_{j,L}$ into Eq.~(\ref{wave6}).

For a scattering problem, the relative wave function is expanded as
\begin{eqnarray}
  \chi_{L}\left(\textbf{R}\right) &=& \sum_{m=1}^{n}C_{m} \frac{\tilde{u}_{L}\left(\boldsymbol{R},\boldsymbol{S}_{m}\right)}{\boldsymbol{R}}Y^L\left(\hat{\boldsymbol{R}}\right),
\end{eqnarray}
with
\begin{equation}\label{QDCSM-vc}
    \tilde{u}_{L}(\boldsymbol{R},\boldsymbol{S}_{m}) = \begin{cases}
        \alpha_{m}u_{L}(\boldsymbol{R},\boldsymbol{S}_{m}), & R \leq R_{C} \\
        \left[h_{L}^{-}(\boldsymbol{k},\boldsymbol{R}) - s_{m}h_{L}^{+}(\boldsymbol{k},\boldsymbol{R})\right]R_{AB}, & R \geq R_{C}
    \end{cases}
\end{equation}
where $u_{L}$ is presented Eq.~(\ref{wave7}), $h^{\pm}_{L}$ is the $L$th spherical Hankel functions, $\boldsymbol{k}$ is the momentum of the relative motion with $\boldsymbol{k}=\sqrt{2\mu E_{ie}}$, $\mu$ is the reduced mass of two hadrons of the open channel, $E_{ie}$ is the incident energy of the relevant open channels, which can be written as $E_{ie}=E_{total}-E_{th}$ where $E_{total}$ denotes the total energy and $E_{th}$ represents the threshold of open channel. $\boldsymbol{R}_{C}$ is a cutoff radius beyond which all the strong interaction can be disregarded. Besides, $\alpha_{m}$ and $s_{m}$ are complex parameters that are determined by the smoothness condition at $ \boldsymbol{R}= \boldsymbol{R}_{C}$ and $C_{m}$ satisfy $\Sigma_{m=1}^{n}C_{m}=1$. After performing the variational procedure, a $L$th partial-wave equation for the scattering problem can be deduced as
\begin{eqnarray}\label{wave11}
  \sum_{m=1}^{n} \mathcal{L}^{L}_{im}C_{m}&=& \mathcal{M}_{i}^{L}(i=0, 1,..., n-1),
\end{eqnarray}

with
\begin{eqnarray}
 \mathcal{L}^{L}_{im}&=& \mathcal{K}_{im}^{L}-\mathcal{K}_{i0}^{L}-\mathcal{K}_{0m}^{L}+\mathcal{K}_{00}^{L},\\
 \mathcal{M}_{i}^{L}&=& \mathcal{K}_{00}^{L}-\mathcal{K}_{i0}^{L},
\end{eqnarray}

and

\begin{equation}
    \begin{split}
        \mathcal{K}_{im}^{L} &= \bigg\langle \hat{\phi}_{A}\hat{\phi}_{B}\frac{\tilde{u}_{L}(\boldsymbol{R}',\boldsymbol{S}_{m})}{\boldsymbol{R}'}Y^{L}(\boldsymbol{R}') \bigg| H-E \\
        &\quad \times \bigg| \mathcal{A}\left[\hat{\phi}_{A}\hat{\phi}_{B}\frac{\tilde{u}_{L}(\boldsymbol{R},\boldsymbol{S}_{m})}{\boldsymbol{R}}Y^{L}(\boldsymbol{R})\right] \bigg\rangle.
    \end{split}
\end{equation}

By solving Eq.(\ref{wave11}), we can obtain the expansion coffefficients $C_{m}$, then the $S$-matrix element $S_{L}$ and the phase shifts $\delta_{L}$ are given by
\begin{eqnarray}
S_{L}=e^{2i\delta_{L}}=\sum_{m=1}^{n}C_{m}s_{m}.
\end{eqnarray}
Finally, the cross-section can be obtained from the scattering phase shifts by the formula,
\begin{eqnarray}\label{wave12}
\sigma_{L} =\frac{4\pi}{k^2}\left(2L+1\right) \sin^{2}\delta_{L}.
\end{eqnarray}

\bibliography{NDbar}

\end{document}